\documentclass[twocolumn,aps,superscriptaddress,showpacs]{revtex4}

\usepackage{amssymb}
\usepackage{amsmath}
\usepackage{mathrsfs}
\usepackage{graphicx}
\usepackage[normalem]{ulem}
\usepackage[dvips]{color}

\begin{document}

\title{Isospin properties of quark matter from a 3-flavor NJL model}
\author{He Liu}
\affiliation{Shanghai Institute of Applied Physics, Chinese Academy
of Sciences, Shanghai 201800, China}
\affiliation{University of Chinese Academy of Sciences, Beijing 100049, China}
\author{Jun Xu\footnote{corresponding author: xujun@sinap.ac.cn}}
\affiliation{Shanghai Institute of Applied Physics, Chinese Academy
of Sciences, Shanghai 201800, China}
\author{Lie-Wen Chen}
\affiliation{Department of Physics and
Astronomy and Shanghai Key Laboratory for Particle Physics and
Cosmology, Shanghai Jiao Tong University, Shanghai 200240, China}
\affiliation{Center of Theoretical Nuclear Physics, National
Laboratory of Heavy Ion Accelerator, Lanzhou 730000, China}
\author{Kai-Jia Sun}
\affiliation{Department of Physics and
Astronomy and Shanghai Key Laboratory for Particle Physics and
Cosmology, Shanghai Jiao Tong University, Shanghai 200240, China}

\date{\today}

\begin{abstract}

We have studied the properties of hot and dense quark matter based
on the 3-flavor Nambu-Jona-Lasinio (NJL) model as well as its
Polyakov-loop extension (pNJL) with scalar-isovector and
vector-isovector couplings. Provided a considerable large isospin
asymmetry or isospin chemical potential, isospin splittings of
constituent mass, chiral phase transition boundary, and critical
point for $u$ and $d$ quarks can be observed for positive isovector
coupling constants but are suppressed for negative ones. The quark
matter symmetry energy decreases with the increasing isovector
coupling constant, and is mostly enhanced in the pNJL model than in
the NJL model. A positive scalar-isovector coupling constant is more
likely to lead to an unstable isospin asymmetric quark matter. The
isovector coupling has been further found to affect particle
fractions as well as the equation of state in hybrid stars. Possible
effects on the isospin properties of quark matter have also been
discussed if the strangeness sector is further broken among the
flavor symmetry.

\end{abstract}

\pacs{21.65.Qr,  
      25.75.Nq,  
      21.65.Ef,  
      26.60.Kp   
      }

\maketitle

\section{Introduction}
\label{introduction}

The phase structure of quantum chromodynamics (QCD) matter in three
dimensions, i.e., the temperature, the baryon chemical potential,
and the isospin, is the holy grail of nuclear physics. So far great
efforts have been made in understanding the QCD phase diagram at
nearly zero isospin. For instance, the transition from the produced
quark-gluon plasma (QGP) to hadronic matter at the top energy of the
Relativistic Heavy Ion Collider (RHIC) or at the Large Hadron
Collider (LHC) is a smooth crossover at nearly zero baryon and
isospin chemical potential based on Lattice QCD (LQCD)
studies~\cite{Ber05,Aok06,Baz12}. Although LQCD suffers from the
fermion sign problem~\cite{Bar86,Kar02,Mur03} at finite baryon
chemical potential, the hadron-quark phase transition can be a
first-order one at large baryon chemical potentials based on studies
from phenomenological theoretical models, e.g., the
Nambu-Jona-Lasinio (NJL) model and its
extensions~\cite{Asa89,Fuk08,Car10,Bra13}. To further explore the
QCD phase structure and search for the signal of the critical point
between the crossover and the first-order transition, experimental
programs such as the Beam-Energy Scan (BES) at RHIC and the
Compressed Baryonic Matter (CBM) at Facilities for Antiproton and
Ion Research (FAIR) were proposed. However, using neutron-rich
heavy-ion beams, the isospin degree of freedom is expected to be
increasingly important at lower collision energies with larger net
baryon densities. The different interactions for $u$ quark and $d$
quark in baryon-rich isospin asymmetric quark matter consisting of
different net numbers of $u$ and $d$ quarks can lead to
isospin-dependent dynamics in the QGP. In addition, the phase
boundary as well as the critical point extracted from these
experiments are actually those at finite isospin in the
3-dimensional QCD phase diagram, since the whole system is a
globally neutron-rich or $d$-quark-rich one.

The isospin physics has been in fact a hot topic in low-energy
nuclear physics as well as nuclear astrophysics in the past 15
years, and the nuclear symmetry energy, the energy excess for
neutron-rich system comparing with isospin symmetric one, is
important in understanding various phenomena in finite nuclei,
nuclear reactions, and compact
stars~\cite{Bar05,Ste05,Lat07,BCK08,Tsa11,Lat12,Hor14}. Generally, a
larger symmetry energy leads to a repulsive (attractive) potential
for particles with negative (positive) isospin, and a stiffer
equation of state (EOS) for isospin asymmetric matter. Besides the
interesting isospin dynamics at RHIC-BES and FAIR-CBM and isospin
dependent QCD phase diagram mentioned above, the quark matter
symmetry energy, the EOS of the isospin asymmetric quark matter, and
their temperature dependence are also interesting topics. The EOS of
hot quark matter is the main ingredient for hydrodynamic calculation
of the QGP evolution, while that of cold quark matter is important
in understanding the properties of quark stars or hybrid stars,
which is probably related to the recently observed massive compact
stars, e.g., PSR J1614-2230 with $\emph{M} =
(1.97\pm0.04)\emph{M}_\odot$~\cite{Dem10} and PSR J0348+0432 with
$\emph{M} = (2.01\pm0.04)\emph{M}_\odot$~\cite{Ant13}. With the
increasing temperature or density in heavy-ion systems or compact
stars, strangeness can also be abundantly produced, and the isospin
dependence of the system is often coupled with the strangeness
sector.

In the present study, we explore the properties of isospin
asymmetric quark matter based on a 3-flavor NJL model, which can
successfully interpret the dynamics of spontaneous breaking of
chiral symmetry in vacuum and its restoration at high temperatures
and baryon chemical potentials, together with its Polyakov-loop
extension (pNJL) which can describe the deconfinement phase
transition. In order to study the isospin physics of quark matter,
we break the SU(3) symmetry of the 3-flavor NJL Lagrangian by
introducing the scalar-isovector and vector-isovector
couplings~\cite{Chu15}, corresponding to different extent of isospin
symmetry breaking in the scalar and vector channels. We will see
that the isospin vector couplings may have dramatic effects on the
isospin dependence of the QCD phase diagram at large isospin
chemical potential as well as the quark matter symmetry energy. We
have also explored the possible effects on the isospin dependence of
the results if the strangeness sector is further broken among the
flavor symmetry.

This paper is organized as follows. In Sec.~\ref{NJL}, we briefly
review the formulism of the 3-flavor NJL model and pNJL model with
isovector couplings. The effects of the isovector couplings on the
phase diagram of isospin asymmetric quark matter and the quark
matter symmetry energy are discussed in Secs.~\ref{diagram} and
\ref{Esym}, and the properties of hybrid stars are discussed in
Sec.~\ref{hybrid stars}. The possible effects from further breaking
of the strangeness sector on the obtained results are discussed in
Sec.~\ref{strangeness}. A summary is given in Sec.~\ref{summary}.

\section{Theoretical model}
\label{NJL}

By introducing the scalar-isovector coupling and the
vector-isovector coupling, the Lagrangian of the 3-flavor NJL model
can be written as
\begin{eqnarray}
\mathcal{L}_{\textrm{NJL}} &=& \bar{q}(i\rlap{\slash}\partial-\hat{m})q
+\frac{G_S}{2}\sum_{a=0}^{8}[(\bar{q}\lambda_aq)^2+(\bar{q}i\gamma_5\lambda_aq)^2]
\notag\\
&+&\frac{G_V}{2}\sum_{a=0}^{8}[(\bar{q}\gamma_\mu\lambda_aq)^2+
(\bar{q}\gamma_5\gamma_\mu\lambda_aq)^2]
\notag\\
&-&K\{\det[\bar{q}(1+\gamma_5)q]+\det[\bar{q}(1-\gamma_5)q]\}
\notag\\
&+&G_{IS}\sum_{a=1}^{3}[(\bar{q}\lambda_aq)^2+(\bar{q}i\gamma_5\lambda_aq)^2]\notag\\
&+&G_{IV}\sum_{a=1}^{3}[(\bar{q}\gamma_\mu\lambda_aq)^2+(\bar{q}\gamma_5\gamma_\mu\lambda_aq)^2],
\end{eqnarray}
where $q$ denotes the quark fields with three flavors, i.e., $u$,
$d$, and $s$, and three colors; $\hat{m}=\text{diag}(m_u, m_d, m_s)$
is the current quark mass matrix in flavor space; $\lambda_a$ are
the flavor SU(3) Gell-Mann matrices with $\lambda_0 = \sqrt{2/3}I$;
$G_S$ and $G_V$ are the strength of the scalar and vector coupling,
respectively; and the $K$ term represents the six-point
Kobayashi-Maskawa-t'Hooft (KMT) interaction that breaks the axial
$U(1)_A$ symmetry. Since the Gell-Mann matrics with $a=1\sim3$ are
identical to the Pauli matrics in $u$ and $d$ space, the last two
terms represent the scalar-isovector and vector-isovector coupling
breaking the SU(3) asymmetry while keeping the isospin symmetry,
with $G_{IS}$ and $G_{IV}$ the corresponding coupling strength. In
the present study, we employ the parameters $m_u = m_d = 3.6$ MeV,
$m_s = 87$ MeV, $G_S\Lambda^2 = 3.6$, $K\Lambda^5 = 8.9$, and the
cutoff value in the momentum integral $\Lambda = 750$ MeV given in
Refs.~\cite{Bra13,Lut92,Bub05}. $G_V$ is set to 0 in the present
study.

In the mean-field approximation, quarks can be taken as
quasiparticles with constituent mass $M_i$ given by the gap equation
as
\begin{eqnarray}
M_i &=&
m_i-2G_S\sigma_i+2K\sigma_j\sigma_k-2G_{IS}\tau_{3i}(\sigma_u-\sigma_d),
\label{mi}
\end{eqnarray}
where $\sigma_i=<\bar{q}_iq_i>$ stands for the quark condensate with
($i$, $j$, $k$) being any permutation of ($u$, $d$, $s$), and
$\tau_{3i}$ is the isospin quantum number of quark, i.e., $\tau_{3u}
= 1$, $\tau_{3d} = -1$, and $\tau_{3s} = 0$. As shown in
Eq.~(\ref{mi}), $\sigma_d$ and $\sigma_s$ contribute to the $u$
quark mass through the KMT interaction as well as the
scalar-isovector coupling, called the flavor mixing~\cite{Fra03,Zha14} in the
constituent quark mass. The constituent quark mass
$M_i$ in vacuum is much larger than the current quark mass $m_i$,
representing the spontaneous chiral symmetry breaking, while at high
densities and/or temperatures $M_i$ becomes approximately the same
as $m_i$, representing the chiral symmetry restoration. In this way,
the quark condensate or the constituent quark mass can serve as an
order parameter for chiral phase transition. In the present study,
the approximate chiral phase transition boundary is taken as where the light
quark condensate is half of that in vacuum~\cite{Fuk08}.

From the mean-field approximation and some algebras based on the
finite-temperature field theory, the thermodynamic potential
$\Omega_\textrm{NJL}$ of quark matter at finite temperature and
quark chemical potential can be expressed as
\begin{eqnarray}\label{omeganjl}
\Omega_{\textrm{NJL}}&=& -2N_c\sum_{i=u,d,s}\int_0^\Lambda\frac{d^3p}{(2\pi)^3}
[E_i+T\ln(1+e^{-\beta(E_i-\tilde{\mu}_i)})
\notag\\
&+&T\ln(1+e^{-\beta(E_i+\tilde{\mu}_i)})]+G_S(\sigma_u^2+\sigma_d^2+\sigma_s^2)
\notag\\
&-&4K\sigma_u\sigma_d\sigma_s+G_V(\rho_u^2+\rho_d^2+\rho_s^2)
\notag\\
&+&G_{IS}(\sigma_u-\sigma_d)^2+G_{IV}(\rho_u-\rho_d)^2.
\end{eqnarray}
In the above, the factor $2N_c$ represents the spin and color
degeneracy, $\beta=1/T$ represents the temperature, $\rho_i$ is the
net quark number density of flavor $i$ ($i=u, d, s$), and
$E_i(p)=\sqrt{p^2 +M_i^2}$ is the single quark energy. The effective
chemical potential $\tilde{\mu}_i$ is defined as
\begin{eqnarray}\label{mui}
\tilde{\mu}_i&=&\mu_i+2G_V\rho_i+2G_{IV}\tau_{3i}(\rho_u-\rho_d),
\end{eqnarray}
with the flavor mixing in $\tilde{\mu}_i$ similar to that in the
constituent quark mass (Eq.~(\ref{mi})). The quark condensate can be
expressed as
\begin{eqnarray}
\sigma_i&=&-2N_c\int_0^\Lambda\frac{d^3p}{(2\pi)^3}\frac{M_i}{E_i}(1-f_i-\bar{f}_i),
\end{eqnarray}
where
\begin{eqnarray}
f_i&=&\frac{1}{1+e^{\beta(E_i-\tilde{\mu}_i)}},
\\
\bar{f_i}&=&\frac{1}{1+e^{\beta(E_i+\tilde{\mu}_i)}},
\end{eqnarray}
are respectively the Fermi distribution functions of quarks and
antiquarks. The net quark number density of the flavor $i$ can be
calculated from $f_i$ and $\bar{f}_i$ via
\begin{eqnarray}
\rho_i=2N_c\int^\Lambda_0(f_i-\bar{f_i})\frac{d^3p}{(2\pi)^3}.
\end{eqnarray}
The above equations are solved self-consistently to obtain the quark
matter properties at a given quark chemical potential and
temperature.

The 3-flavor NJL model briefly reviewed above is effective in
describing chiral phase transition but fails to get a deconfinement
transition. The Polyakov loop, which was inspired by the
strong-coupling analyses~\cite{Goc85, Ilg85, Dig01, Moc04}, has
later been incorporated into the NJL model~\cite{Fuk04,Rat06} in
order to compensate effectively the gluon contribution, and it can
serve as an order parameter for deconfinement phase
transition~\cite{Fuk04,Fuk11}.

The thermodynamic potential $\Omega_\textrm{pNJL}$ of the 3-flavor
pNJL model at finite temperature and quark chemical potential can be
expressed as
\begin{eqnarray}\label{omegapnjl}
\Omega_{\textrm{pNJL}} &=&\mathcal{U}(\Phi,\bar{\Phi},T)-2N_c\sum_i\int_0^\Lambda\frac{d^3p}{(2\pi)^3}E_i
\notag\\
&-&2T\sum_i\int\frac{d^3p}{(2\pi)^3}[\ln(1+e^{-3\beta(E_i-\tilde{\mu}_i)}
\notag\\
&+&3\Phi e^{-\beta(E_i-\tilde{\mu}_i)}
+3\bar{\Phi}e^{-2\beta(E_i-\tilde{\mu}_i)})
\notag\\
&+&\ln(1+e^{-3\beta(E_i+\tilde{\mu}_i)}
+3\bar{\Phi} e^{-\beta(E_i+\tilde{\mu}_i)}
\notag\\
&+&3\Phi e^{-2\beta(E_i+\tilde{\mu}_i)})]
+G_S(\sigma_u^2+\sigma_d^2+\sigma_s^2)
\notag\\
&-&4K\sigma_u\sigma_d\sigma_s +G_V(\rho_u^2+\rho_d^2+\rho_s^2)
\notag\\
&+&G_{IS}(\sigma_u-\sigma_d)^2
+G_{IV}(\rho_u-\rho_d)^2,
\end{eqnarray}
where the form of the temperature-dependent effective potential
$\mathcal{U}(\Phi, \bar{\Phi}, T)$ as a function of the Polyakov loop $\Phi$ and $\bar{\Phi}$ is taken from Ref.~\cite{Fuk08}
as
\begin{eqnarray}
\mathcal{U}(\Phi,\bar{\Phi},T) &=& -b \cdot
T\{54e^{-a/T}\Phi\bar{\Phi} +\ln[1-6\Phi\bar{\Phi}
\notag\\
&-&3(\Phi\bar{\Phi})^2+4(\Phi^3+\bar{\Phi}^3)]\}.
\end{eqnarray}
The parameters
$a=664$ MeV and $b=0.015\Lambda^3$ are determined by the condition
that the first-order phase transition in the pure gluodynamics takes
place at $T = 270$ MeV, and the simultaneous crossover of chiral
restoration and deconfinement phase transition occurs around $T
\approx 200$ MeV~\cite{Fuk08}. The second integral in Eq.~(\ref{omegapnjl}) is finite thus without the ultraviolet cutoff, different from the NJL model. In order
to get the minimum of the thermodynamic potential
$\Omega_\textrm{pNJL}$, the following five equations are solved
\begin{eqnarray}
\frac{\partial\Omega_{\textrm{pNJL}}}{\partial\sigma_u}
=\frac{\partial\Omega_{\textrm{pNJL}}}{\partial\sigma_d}
=\frac{\partial\Omega_{\textrm{pNJL}}}{\partial\sigma_s}
=\frac{\partial\Omega_{\textrm{pNJL}}}{\partial\Phi}
=\frac{\partial\Omega_{\textrm{pNJL}}}{\partial\bar{\Phi}} =0,
\notag
\end{eqnarray}
leading to the values of $\sigma_u$, $\sigma_d$, $\sigma_s$, $\Phi$,
and $\bar{\Phi}$ in the pNJL model. The approximate deconfinement
phase transition boundary is taken as where the Polyakov loop $\Phi$
is equal to $1/2$~\cite{Fuk08}.

Starting from the thermodynamic potential, the energy density of the
system can be obtained from the thermodynamical relation
\begin{eqnarray}
\varepsilon&=&\Omega+\beta\frac{\partial}{\partial\beta}\Omega+\sum_i\mu_i\rho_i.
\end{eqnarray}
Accordingly, the energy density from the NJL model can
be written as
\begin{eqnarray}\label{epsilon}
\varepsilon_{\textrm{NJL}}&=&-2N_c\sum_{i=u,d,s}\int_0^\Lambda\frac{d^3p}{(2\pi)^3}
E_i(1-f_i-\bar{f}_i)
\notag\\
&-&\sum_{i=u,d,s}(\tilde{\mu}_i-\mu_i)\rho_i+G_S(\sigma_u^2+\sigma_d^2+\sigma_s^2)
\notag\\
&-&4K\sigma_u\sigma_d\sigma_s+G_V(\rho_u^2+\rho_d^2+\rho_s^2)
\notag\\
&+&G_{IS}(\sigma_u-\sigma_d)^2+G_{IV}(\rho_u-\rho_d)^2-\varepsilon_0.
\end{eqnarray}
In the above expression, $\varepsilon_0$ is introduced to ensure
$\varepsilon_{\textrm{NJL}} = 0$ in vacuum. Similarly, the energy
density from the pNJL model can be expressed as
\begin{eqnarray}\label{epsilon_pNJL}
\varepsilon_{\textrm{pNJL}}&=&
54abe^{-a/T}\Phi\bar{\Phi}+2N_c\sum_{i=u,d,s}\int_\Lambda^\infty\frac{d^3p}{(2\pi)^3}
E_i
\notag\\
&-&2N_c\sum_{i=u,d,s}\int\frac{d^3p}{(2\pi)^3}
E_i(1-F_i-\bar{F}_i)\notag\\
&-&\sum_{i=u,d,s}(\tilde{\mu}_i-\mu_i)\rho_i+G_S(\sigma_u^2+\sigma_d^2+\sigma_s^2)
\notag\\
&-&4K\sigma_u\sigma_d\sigma_s+G_V(\rho_u^2+\rho_d^2+\rho_s^2)
\notag\\
&+&G_{IS}(\sigma_u-\sigma_d)^2+G_{IV}(\rho_u-\rho_d)^2-\varepsilon_0,
\end{eqnarray}
where
\begin{equation}
F_i = \frac{1+2\bar\Phi\xi_i+\Phi\xi_i^2}{1+ 3\bar\Phi\xi_i
+3\Phi\xi_i^2+\xi_i^3}
\end{equation}
and
\begin{equation}
\bar F_i=
\frac{1+2\Phi{\xi^\prime_i}+\bar\Phi{\xi^\prime_i}^2}{1+3\Phi{\xi^\prime_i}+3\bar\Phi{\xi^\prime_i}^2+{\xi^\prime_i}^3}
\end{equation}
are the effective phase-space distribution functions for quarks and
antiquarks in the pNJL model with $\xi_i =
e^{(E_i-\tilde{\mu}_i)/T}$ and $\xi_i^\prime =
e^{(E_i+\tilde{\mu}_i)/T}$. One expects that the different effective
phase-space distribution functions for quarks and antiquarks in the
pNJL model may lead to different temperature effects on the
thermodynamical quantities from the NJL model. The pressure for cold
quark matter can be calculated from
\begin{eqnarray}
P=\sum_{i=u, d,
s}\mu_i\rho_i-\varepsilon_{\textrm{NJL}},\label{pressure}
\end{eqnarray}
which will be used in the study of compact stars.

\section{Isospin dependence of phase diagram}
\label{diagram}

Most of our knowledge on the QCD phase diagram are restricted to
zero isospin chemical potential $\mu_I=0$. In relativistic heavy-ion
collisions using neutron-rich nucleus beams, the hadron-quark phase
transition is related to the phase diagram at nonzero $\mu_I$ or isospin asymmetry $\delta$. In
terms of different chemical potential $\mu_u$ for $u$ quarks and
$\mu_d$ for $d$ quarks, the baryon chemical potential $\mu_B$ and the
isospin chemical potential $\mu_I$ can be expressed respectively as
\begin{eqnarray}
\frac{\mu_B}{3} = \frac{\mu_u+\mu_d}{2} = \mu,~~\qquad \mu_I =
\frac{\mu_u-\mu_d}{2}.
\end{eqnarray}
The isospin asymmetry $\delta$ in the quark phase can be defined as~\cite{Tor11}
\begin{eqnarray}
\delta=3\frac{\rho_d-\rho_u}{\rho_d+\rho_u},
\end{eqnarray}
where the $\rho_u$ and $\rho_d$ are the net quark number densities
for $u$ and $d$ quarks, respectively. The above definitions of the
isospin chemical potential $\mu_I$ and the isospin asymmetry
$\delta$ can be consistently related to those in nuclear matter
\begin{eqnarray}
\mu_I &=& \frac{\mu_p-\mu_n}{2}, \\
\delta &=& \frac{\rho_n-\rho_p}{\rho_n+\rho_p},
\end{eqnarray}
with $\mu_n$ and $\mu_p$ being the neutron and proton chemical
potentials and $\rho_n$ and $\rho_p$ the corresponding number
densities. By assuming that the ratio of electric/baryon charge, or
equivalently the isospin asymmetry $\delta$ if the net strange quark
number is zero, is conserved in relativistic heavy-ion collisions,
the isospin asymmetry in the quark phase produced in central Au+Au
collisions is thus
\begin{equation}
\delta = \frac{N-Z}{N+Z}=0.198,
\end{equation}
with $N=118$ and $Z=79$ being the neutron and proton numbers for Au
nucleus, respectively. Due to the larger degeneracy of quarks than
nucleons, the isospin chemical potential is much smaller in the
quark phase than in the nucleon phase at the same isospin asymmetry,
especially at lower temperatures. Similar to
Refs.~\cite{Tou03,Fra03,Zha14}, we will study the QCD phase diagram
at a fixed isospin chemical potential $\mu_I=-30$ MeV in addition to
that at a fixed isospin asymmetry $\delta=0.198$ in the following,
while the net strange quark density $\rho_s$ is assumed to be zero
in both cases. We must note here that an isospin chemical potential
as large as $\mu_I=-30$ MeV can not be reached so far in heavy-ion
experiments according to our best knowledge, but it is always of
theoretical interest to explore the QCD phase diagram at larger
$\mu_I$. We will also discuss the influence of the scalar-isovector
and vector-isovector interactions on the isospin dependence of the
QCD phase diagram. For the ease of discussions, we define
$R_{IS}=G_{IS}/G_S$ and $R_{IV}=G_{IV}/G_S$ as the reduced strength
of the scalar-isovector and vector-isovector coupling. Since the NJL
model can be considered as an effective field theory, $R_{IS}$ and
$R_{IV}$ are treated as free parameters in the following studies.

\begin{figure}[tbh]
\includegraphics[scale=0.9]{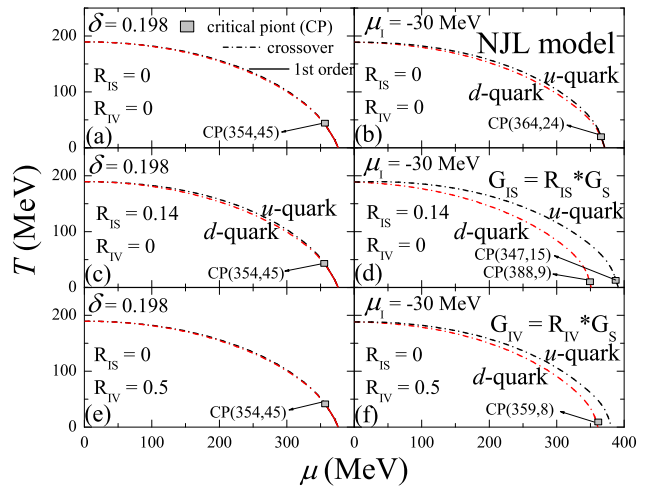}
\caption{(color online) Phase diagram in the $\mu - T$ plane of quark matter from the 3-flavor
NJL model with
various scalar-isovector and vector-isovector coupling constants at a fixed isospin asymmetry $\delta=0.198$ (left) as in Au+Au collisions or at a fixed isospin chemical potential
$\mu_I=-30$ MeV (right). The dash-dotted curves represent the approximate phase
boundaries of a crossover transition while the solid lines are those
of a first-order transition, with a critical point appearing in
between.} \label{gis}
\end{figure}

We begin our discussion with the phase diagram from the 3-flavor NJL
model. Figure~\ref{gis} displays the phase diagram in the $\mu - T$
plane for various isovector coupling constants at a fixed isospin
asymmetry or isospin chemical potential, and with zero net strange
quark density. The left panels represent the phase diagram at the
isospin asymmetry $\delta=0.198$, and the critical point for the
chiral phase transition is ($\mu=354$ MeV, $T=45$ MeV), representing
the transition of the light quark condensate from a smooth change to
a sudden jump at the phase boundary. It is seen that the QCD phase
diagram is almost insensitive to the isovector couplings, as a
result of the small isospin asymmetry or isospin chemical potential
reached in Au+Au collisions. At a larger isospin chemical potential
$\mu_I=-30$ MeV without isovector couplings (Panel (b)), the chiral
phase transitions of $u$ quark and $d$ quark still share the similar
boundary and the critical point, as a result of the flavor-mixing
effect due to the six-point interaction which refers to the axial
$U(1)_A$ symmetry anomaly~\cite{Fra03}. With the increasing
scalar-isovector coupling constant, the phase boundaries as well as
the critical points of $u$ and $d$ quarks become to separate and
their difference reaches the maximum around $G_{IS}=0.14G_S$, as
shown in Panel (d), where the temperatures of the two critical
points are also lower compared to the case without isovector
couplings. The isospin splitting of the $u$ and $d$ quark chiral
phase transition has also been observed in
Refs.~\cite{Tou03,Fra03,Zha14}. Further increase of $R_{IS}$ leads
to a negative $d$ quark constituent mass near the phase boundary and
will be discussed later. We find that for negative values of
$R_{IS}$ there is no separation of the $u$ and $d$ quark chiral
phase transition and the critical point is similar to that without
isovector couplings. We further display the effects of the
vector-isovector coupling on the phase diagram in Panel (f), and
observe similar effects as those from the scalar-isovector coupling,
i.e., the isospin splitting of the chiral phase transition boundary
is observed for positive $R_{IV}$, while for negative $R_{IV}$ the
phase diagram is the same as that without isovector couplings. For
$R_{IV}$ larger than 0.5, the $d$ quark chemical potential near the
phase boundary is comparable to the cutoff value $\Lambda$ in the
momentum integral, i.e., leading to the invalidity of the model. For
positive $R_{IS}$ or $R_{IV}$, it is also observed that the critical
point for $d$ quarks is always at a slightly smaller chemical
potential compared with that for $u$ quarks.

\begin{figure}[tbh]
\includegraphics[scale=0.85, bb = 0 0 273 208]{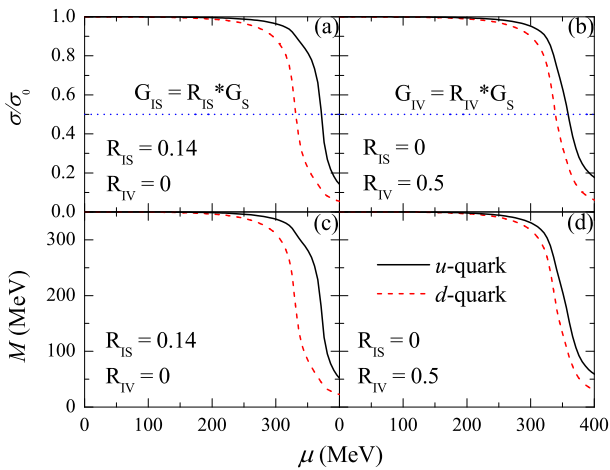}
\caption{(color online) The reduced quark condensates (upper panels)
and the constituent masses (lower panels) of $u$ and $d$ quarks as a
function of quark chemical potential at $T=50$ MeV and
$\mu_I=-30$ MeV with typical strength of
scalar-isovector and vector-isovector couplings. $\sigma_0$ is the
quark condensate in vacuum.} \label{quar}
\end{figure}

To understand the phase diagrams in Fig.~\ref{gis}, we show in
Fig.~\ref{quar} the reduced light quark condensate (upper panels)
and the corresponding constituent quark mass (lower panels) at a
typical temperature of $T=50$ MeV and isospin chemical potential
$\mu_I=-30$ MeV. The $u$ quark condensate has a
larger magnitude than the $d$ quark condensate, resulting in a
separate phase boundary at the chemical potential where the reduced
quark condensate is 1/2. The different $u$ and $d$ quark condensates
lead to their different constituent quark masses according to
Eq.~(\ref{mi}), resulting in a larger $u$ quark than $d$ quark mass.
The isospin splitting of the constituent quark mass is important in
isospin dynamics of relativistic heavy-ion collisions. For instance,
if $u$ and $d$ quarks are affected by the same potential, $d$ quarks
will propagate faster. Positive values of $R_{IS}$ lead to the
enhancement of the isospin splitting of quark condensate as well as
the constituent quark mass through the positive feedback mechanism,
as can be seen from Eq.~(\ref{mi}) that a larger difference between
$\sigma_u$ and $\sigma_d$ leads to a larger difference between $M_u$
and $M_d$, given that the quark condensate is negative. On the other
hand, negative values of $R_{IS}$ reduce the isospin splitting of
quark condensate as well as that of the constituent quark mass
through the negative feedback mechanism, leading to eventually the
same phase boundary for $u$ and $d$ quarks. The values of $R_{IS}$
greater than 0.14 lead to a too large isospin splitting and thus a
negative $d$ quark mass near the phase boundary. Similar mechanism
applies to the vector-isovector coupling, but with reduced secondary
effect.

\begin{figure}[tbh]
\includegraphics[scale=0.9]{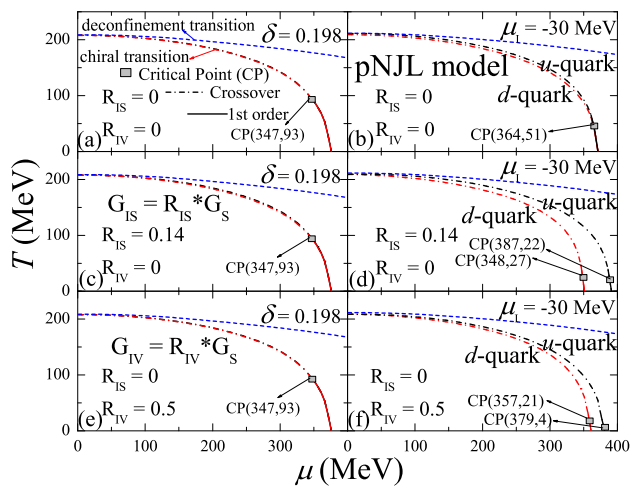}
\vspace{-0.5cm}\caption{(color online) Similar to Fig.~\ref{gis} but
from the 3-flavor pNJL model. The dashed lines represent the
approximate phase boundary of a crossover deconfinement transition.}
\label{pgisv}
\end{figure}

The effect of the Polyakov loop on the phase diagram in the
$\mu$-$T$ plane is displayed in Fig.~\ref{pgisv}. As found in many
other works~\cite{Fuk08,Car10}, the critical point moves to a higher
temperature and a slightly smaller quark chemical potential
($\mu=347$ MeV, $T=93$ MeV) from the pNJL model at the isospin asymmetry $\delta=0.198$ without isovecter
couplings, as exhibited in the left panels. The isospin effect on the QCD phase diagram with the Polyakov loop is still small in the quark system produced from Au+Au collisions. With isovector couplings at
$\mu_I=-30$ MeV, the isospin splitting of the $u$
and $d$ quark chiral phase boundaries as well as their critical
points is observed, qualitatively similar to the results from the
NJL model. On the other hand, the temperatures of the critical
points are also mostly increased in the pNJL model compared with
those in Fig.~\ref{gis}. The approximate phase boundary of
deconfinement transition is plotted, which is always a crossover in
the pNJL model and seems to be mostly independent of the isovector
couplings. The region between the dashed line and the dash-dotted
line was referred to as the quarkyonic phase in
Refs.~\cite{Fuk08,Sas07,McL07,Glo08}, while a latest study found
that the chiral symmetry can be broken in quarkyonic matter in an
inhomogeneous way, which is called the quarkyonic chiral
spirals~\cite{Koj10a,Koj10b}.

To summarize, the isospin effect on the QCD phase diagram is small with the small isospin asymmetry or isospin chemical potential reached in Au+Au collisions, while the isospin splitting of $u$ and $d$ quark chiral phase transition boundaries and critical points becomes considerable at a fixed larger isospin chemical potential $\mu_I=-30$ MeV. Since the quark condensate is the order parameter for
chiral phase transition, the scalar-isovector coupling has the
direct effect on the isospin splitting of $u$ and $d$ quark chiral
phase transition boundaries and critical points, while the
vector-isovector coupling has the similar but secondary effect. This
can be understood from Eq.~(\ref{mi}). The Polyakov loop doesn't
affect the isospin dependence of the phase diagram but moves the
critical point to higher temperatures.

\section{Quark matter Symmetry energy}
\label{Esym}

The symmetry energy in nucleonic system, denoting the energy
difference between isospin asymmetric and symmetric nuclear matter,
has been a hot topic so
far~\cite{Bar05,Ste05,Lat07,BCK08,Tsa11,Lat12,Hor14}. Since the
quark matter in relativistic heavy-ion collisions at RHIC-BES or
FAIR-CBM energies as well as in compact stars is isospin asymmetric,
the quark matter symmetry energy is also an important quantity
affecting the EOS of the system. It has been shown that the quark
matter symmetry energy is important in understanding the properties
of quark stars and explaining the observed two-solar-mass compact
stars based on a confined-isospin-density-dependent-mass
model~\cite{Chu14}. On the other hand, the importance of the quark
matter symmetry energy is not restricted to the fact that it is a
piece of the EOS of the system, but it is related to the isospin
splitting of $u$ and $d$ quark constituent mass as well as their
different potentials, with the latter splitting as $\pm
G_{IV}(\rho_u-\rho_d)$ in isospin asymmetric quark matter, through
respectively the scalar-isovector and the vector-isovector coupling.
This is similar to the case of nuclear matter symmetry energy which
is related to the isospin splitting of neutron and proton in-medium
effective mass as well as their mean-field
potentials~\cite{BCK08,BAL13}.

Generally, the binding energy of quark matter consisting of $u$,
$d$, and $s$ quarks can be expanded in isospin asymmetry as
\begin{eqnarray}\label{E}
E(\rho_B, \delta, \rho_s)=E_0(\rho_B, \rho_s)+ E_{sym}(\rho_B,
\rho_s)\delta^2+\vartheta(\delta^4).
\end{eqnarray}
In the above, $E_0(\rho_B, \rho_s) = E(\rho_B, \delta = 0, \rho_s)$
is the binding energy per baryon number in the 3-flavor quark
matter with equal number of $u$ and $d$ quarks and net $s$ quark
number density $\rho_s$, and $\rho_B = (\rho_u+\rho_d+\rho_s)/3$ is the
net baryon number density. The quark matter symmetry
energy $E_{sym}(\rho_B, \rho_s)$, standing as the second-order
coefficient in the expansion of the isospin asymmetry, is expressed
by definition as
\begin{eqnarray}
E_{sym}(\rho_B, \rho_s)&=&\frac{1}{2!}\frac{\partial^2E(\rho_B, \delta, \rho_s)}{\partial\delta^2}\mid_{\delta=0}.
\end{eqnarray}
Note that the definition of the quark matter symmetry energy in the
present study has been generalized to finite-temperature systems
containing both quarks and antiquarks. Neglecting the contribution
from higher-order terms in Eq.~(\ref{E}) for small $\delta$, the quark matter symmetry
energy can also be calculated approximately from
\begin{eqnarray}
E_{sym}(\rho_B, \rho_s) \approx \frac{E(\rho_B, \delta,
\rho_s)-E(\rho_B, \delta=0, \rho_s)}{\delta^2},
\end{eqnarray}
where $\delta=0.05$ is used in the calculation. One sees that
$E_{sym}(\rho_B, \rho_s)$ depends on both $\rho_B$ and $\rho_s$,
corresponding to the interplay between the isospin and the
strangeness sectors.

\begin{figure}[tbh]
\includegraphics[scale=0.7]{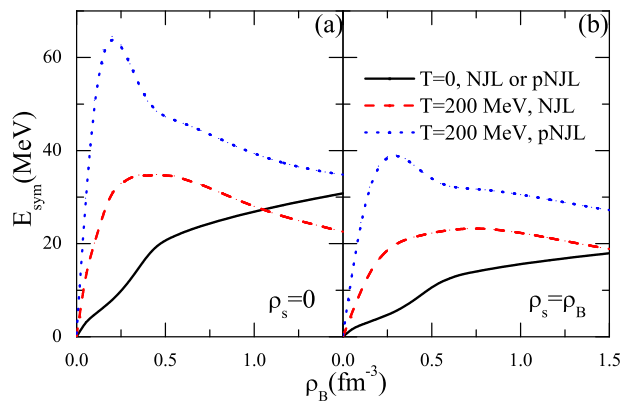}
\caption{(color online) Quark matter symmetry energy from the
3-flavor NJL and pNJL model with strange quark density
$\rho_s=0$ (left) and $\rho_s=\rho_B$ (right) at $T=0$ and $T=200$
MeV without isovector couplings. } \label{ssym}
\end{figure}

We begin the discussion with the quark matter symmetry energy
$E_{sym}$ by comparing that from the NJL model and the pNJL model
without isovector couplings, as displayed in Fig.~\ref{ssym}, with
results from $\rho_s=0$ in the left panel and $\rho_s=\rho_B$ in the
right panel. At zero temperature when the pNJL model reduces to the
NJL model, $E_{sym}$ increases monotonically with increasing
$\rho_B$, while at $T=200$ MeV it first increases then slightly
decreases with increasing $\rho_B$. The density and temperature
dependence of $E_{sym}$ are not so simple, but we found that there
are always dramatic changes of the quantities near the chiral
transition phase boundary. The quark matter symmetry energy is much
enhanced in the pNJL model at $T=200$ MeV, especially at lower
densities, compared to that in the NJL model. The larger quark
matter symmetry energy in the pNJL model than in the NJL model is
mainly due to their different kinetic energy contributions, as a
result of larger isospin splitting of $u$ and $d$ quark constituent
masses as well as a more diffusive phase-space distribution function
in the pNJL model. On the other hand, $E_{sym}$ is reduced in the presence of
strange quarks for both NJL model and pNJL model, and the density is
rescaled, i.e., the whole curve moves to the high-density side since
in this case we have $\rho_B=\frac{1}{2}(\rho_u+\rho_d)$ instead of
$\rho_B=\frac{1}{3}(\rho_u+\rho_d)$.

\begin{figure}[tbh]
\includegraphics[scale=0.85]{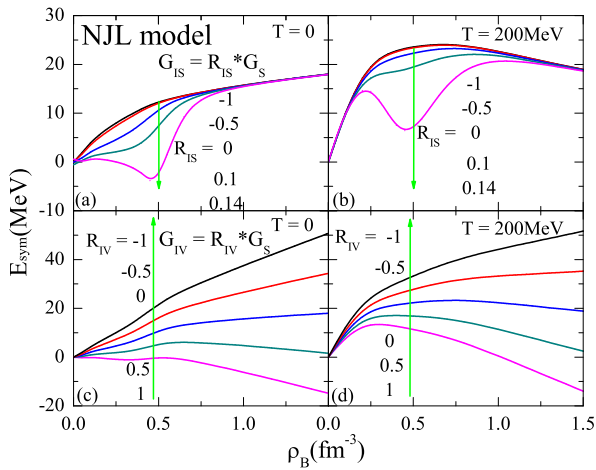}
\caption{(color online) Quark matter symmetry energy for different
scalar-isovector (upper panels) and vector-isovector (lower panels)
coupling constants at $T=0$ (left panels) and $200$ MeV (right
panels) from the 3-flavor NJL model with $\rho_s=\rho_B$. } \label{sym}
\end{figure}

The quark matter symmetry energy for various isovector coupling
constants from the 3-flavor NJL model with $\rho_s=\rho_B$ is displayed in
Fig.~\ref{sym}. Qualitatively, $E_{sym}$ decreases with increasing
constant of both the scalar-isovector and vector-isovector
couplings. It is seen that $E_{sym}$ is sensitive to the
scalar-isovector coupling only at intermediate densities when the
difference between the $u$ and $d$ quark condensate is large, and it
exhibits a strong decrease for $R_{IS}=0.14$ when that difference is
largest. In addition, $E_{sym}$ decreases linearly with increasing
$R_{IV}$. The sensitivity of $E_{sym}$ on the scalar-isovector and
vector-isovector couplings can be understood respectively from
Eq.~(\ref{epsilon}) that the first flavor summation overwhelms the
$G_{IS}(\sigma_u-\sigma_d)^2$ term and the second flavor summation
overwhelms the $G_{IV}(\rho_u-\rho_d)^2$ term, based on some
algebras. The temperature effect on $E_{sym}$ is larger at lower
$\rho_B$ while smaller at higher $\rho_B$.

\begin{figure}[tbh]
\includegraphics[scale=0.85]{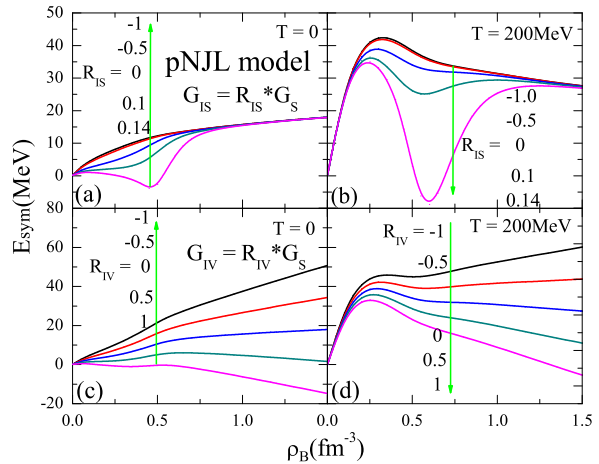}
\caption{(color online) Same as Fig.~\ref{sym} but from the 3-flavor
pNJL model. } \label{psym}
\end{figure}

Figure~\ref{psym} displays the quark matter symmetry energy from the
3-flavor pNJL model with $\rho_s=\rho_B$. Again, the dependence of the scalar-isovector
and vector-isovector coupling on $E_{sym}$ is qualitatively similar
to that from the NJL model. On the other hand, except for the case
with a positive $R_{IS}$, $E_{sym}$ is mostly enhanced in the pNJL
model compared with that in the NJL model, especially at lower
baryon densities.

To summarize, the vector-isovector coupling has the direct effect on
the quark matter symmetry energy, while the scalar-isovector
coupling has the secondary effect. This can be understood from
Eq.~(\ref{epsilon}). The Polyakov loop mostly enhances the quark
matter symmetry energy, compared with that from the NJL model, and
this is mainly due to the different kinetic contributions.

We would also like to emphasize here that we take the
scalar-isovector coupling $G_{IS}$ and the vector-isovector coupling
$G_{IV}$ as the two sources of the quark matter symmetry energy, in
addition to the naive kinetic contribution. On the other hand,
$G_{IS}$ and $G_{IV}$ can lead to other isospin effects in addition
to the quark matter symmetry energy, such as the $u$ and $d$ quark
constituent mass splitting. Once we know the quark matter symmetry
energy and some other microscopic properties of quark matter, we can
have a global picture of quark interactions. This could be achieved
by comparing results from transport model simulations using the NJL
Lagrangian with available experimental data in the near future.

\section{Applications to hybrid stars}
\label{hybrid stars}

The quark matter symmetry energy obtained above can be important in
understanding properties of compact stars with quark degree of
freedom. It has been shown~\cite{Chu15} that the vector-isovector
coupling has considerable effects on the EOS of strange quark star
matter, a $\beta$-equilibrium and charge-neutral system containing
$u$, $d$, and $s$ quarks as well as leptons. Here we apply the
present model to hybrid stars, with a quark core at high densities,
a mixed phase of quarks and hadrons at moderate densities, and a
hadronic phase at low densities. The possible appearance of hyperons
is neglected and in this work we mainly focus on the quark matter
effects on the properties of hybrid stars.

In the high-density quark phase, the system consists of a mixture of
quarks ($u$, $d$, and $s$) and leptons ($e$ and $\mu$) at charge
neutrality and $\beta$-equilibrium condition, i.e.,
\begin{equation}
\frac{2}{3}\rho_u-\frac{1}{3}(\rho_d+\rho_s)-\rho_e-\rho_\mu=0
\end{equation}
and
\begin{equation}
\mu_i=\mu_Bb_i-\mu_cq_i,
\end{equation}
with $\mu_B$ and $\mu_c$ being the baryon and charge chemical
potential, and $q_i$ and $b_i$ being the charge and baryon number of
particle species $i$, respectively. For quarks, the energy density
($\varepsilon_Q$) and the pressure ($P_Q$) can be obtained based on
the NJL model from Eqs.~(\ref{epsilon}) and (\ref{pressure}),
respectively. For leptons, we take both electrons and muons as free
Fermi gas with their masses $m_e=0$ and $m_\mu=106$ MeV,
respectively, and their energy density and pressure can be expressed
as
\begin{eqnarray}
\varepsilon_L&=&\sum_{i=e,
\mu}\frac{1}{\pi^2}\int_0^{k_F^i}\sqrt{k^2+m_i^2}k^2dk,
\\
P_L&=&\sum_{i=e, \mu}\mu_i\rho_i-\varepsilon_L.
\end{eqnarray}
The total energy density and pressure of the high-density quark
phase in hybrid stars including the contributions from both quarks
and leptons are given by
\begin{eqnarray}
\varepsilon^Q&=&\varepsilon_Q+\varepsilon_L,
\\
P^Q&=&P_Q+P_L.
\end{eqnarray}

In the low-density hadronic phase, an isospin- and
momentum-dependent effective nuclear interaction is used to describe
the $\beta$-equilibrium and charge-neutral neutron star matter, with
the single-particle potential written as~\cite{Das03,Che05}
\begin{eqnarray}
U_\tau(\rho ,\delta ,\vec{p}) &=&A_{u}\frac{\rho _{-\tau }}{\rho _{0}}%
+A_{l}\frac{\rho _{\tau }}{\rho _{0}}  \notag \\
&+&B\left(\frac{\rho }{\rho _{0}}\right)^{\sigma }(1-x\delta ^{2})-4\tau x\frac{B}{%
\sigma +1}\frac{\rho ^{\sigma -1}}{\rho _{0}^{\sigma }}\delta \rho
_{-\tau }
\notag \\
&+&\frac{2C_l}{\rho _{0}}\int d^{3}p^{\prime }\frac{f_{\tau }(%
\vec{r},\vec{p}^{\prime })}{1+(\vec{p}-\vec{p}^{\prime
})^{2}/\Lambda ^{2}}
\notag \\
&+&\frac{2C_u}{\rho _{0}}\int d^{3}p^{\prime }\frac{f_{-\tau }(%
\vec{r},\vec{p}^{\prime })}{1+(\vec{p}-\vec{p}^{\prime
})^{2}/\Lambda ^{2}}. \label{MDIU}
\end{eqnarray}%
In the above,  $\tau=1(-1)$ for neutrons (protons) is the isospin
index, $\rho_n$ and $\rho_p$ are number densities of neutrons and
protons, respectively, the isospin asymmetry $\delta$ is defined as
$\delta=(\rho_n-\rho_p)/\rho$, with $\rho=\rho_n+\rho_p$ being the
total number density, and $f_{\tau }(\vec{r},\vec{p})$ is the
nucleon phase-space distribution function. The seven parameters
($A_l$, $A_u$, $B$, $C_l$, $C_u$, $\Lambda$, $\sigma$) are fitted by
empirical constraints of nuclear matter properties at the saturation
density, and their detailed values can be found in
Ref.~\cite{Che05}. In the present study we set $x=0$ corresponding
to a moderately stiff nuclear matter symmetry energy. The chemical
potential of neutrons and protons can be calculated from
\begin{equation}
\mu_\tau = \sqrt{m^2+{p_\tau^F}^2} + U_\tau(\rho,\delta,p_\tau^F),
\end{equation}
with $m$ the nucleon mass and $p_\tau^F=(3\pi^2\rho_\tau)^{1/3}$ the
Fermi momentum. The total energy density and pressure of the
low-density hadronic phase in hybrid stars including the
contributions from both nucleons and leptons are given by
\begin{eqnarray}
\varepsilon^H&=&\varepsilon_H+\varepsilon_L,
\\
P^H&=&P_H+P_L,
\end{eqnarray}
where the detailed expressions for the energy density
$\varepsilon_H$ and pressure $P_H$ of nuclear matter can be found in
Ref.~\cite{Jxu09J}.

At moderate densities of hybrid stars, the Gibbs construction
method~\cite{Gle92,Gle01} is applied to construct the hadron-quark
mixed phase, with the $\beta$-equilibrium, the baryon number
conservation, and the charge neutrality conditions respectively
expressed as
\begin{eqnarray}
\mu_i&=&\mu_Bb_i-\mu_cq_i,
\\
\rho_B&=&(1-Y)(\rho_n+\rho_p)+\frac{Y}{3}(\rho_u+\rho_d+\rho_s),
\label{rhob}\\
0&=&(1-Y)\rho_p+\frac{Y}{3}(2\rho_u-\rho_d-\rho_s)-\rho_e-\rho_\mu,
\end{eqnarray}
where $Y$ is the baryon number fraction of the quark phase. The
total energy density and pressure of the mixed phase are calculated
according to
\begin{eqnarray}
\varepsilon^M&=&(1-Y)\varepsilon_H+Y\varepsilon_Q+\varepsilon_L,
\\
P^M&=&(1-Y)P_H+YP_Q+P_L.
\end{eqnarray}

The core-crust transition density as well as the crust EOS in hybrid
stars are also treated properly according to
Refs.~\cite{Jxu09C,Jxu09J}. The whole EOS from low to high densities
is used to study the mass-radius relation of hybrid stars through
the Tolman-Oppenheimer-Volkoff (TOV) equations~\cite{Opp39}.
Technical details of the study can be found in Ref.~\cite{Jxu10}
except that the quark matter is now described by the 3-flavor NJL
model.

As described above, compact stars are different systems compared
with heavy-ion collisions. The latter has a short lift time, a
higher temperature, and zero net strangeness, while the former is a
stable cold system with particle species determined from the
$\beta$-equilibrium and charge neutrality condition. Investigating
the two systems could be helpful in understanding the quark
interaction from different points of view.

\begin{figure}[tbh]
\includegraphics[scale=0.8,bb = 0 0 329 230]{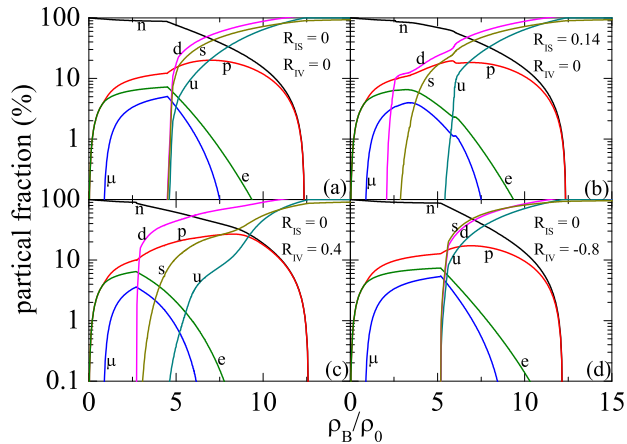}
\caption{(color online) Particle fractions of neutrons ($n$),
protons ($p$), electrons ($e$), muons ($\mu$), $u$ quarks ($u$), $d$
quarks ($d$), and $s$ quarks ($s$) in hybrid star matter with quark
matter properties described by the 3-flavor NJL model from various
isovector coupling constants. } \label{frac}
\end{figure}

Before discussing the EOS, we first display the particle fractions
in hybrid stars in the whole density range in Fig.~\ref{frac}. With
various isovector coupling strength, the hadron-quark mixed phase
appears around $3 \sim 5$ $\rho_0$ and disappears at
$\rho_B>10\rho_0$, where leptons are largely suppressed. For
positive isovector coupling constants, quarks of different flavors
appear at different densities and their fractions are quite
different, especially for the positive vector-isovector coupling
constants which directly affects the quark chemical potential via
Eq.~(\ref{mui}). For negative isovector coupling constants, quarks
of different flavors appear at almost the same density, similar to
the case without isovector coupling, as a result of negative isospin
feedback mechanism discussed in Sec.~\ref{diagram}. It is also
observed that quarks generally appear at higher densities for
negative isovector coupling constants compared with positive ones.

\begin{figure}[tbh]
\includegraphics[scale=0.8]{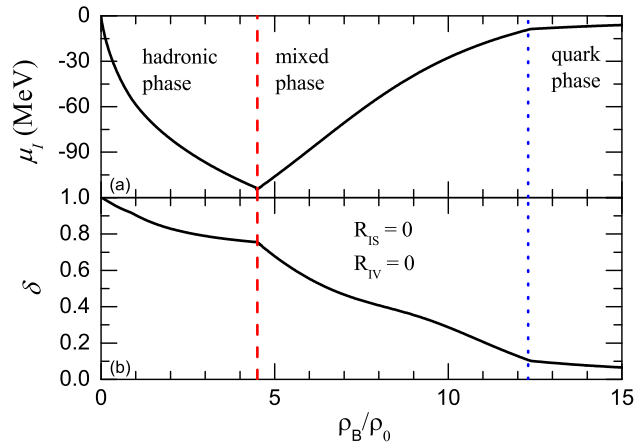}
\caption{(color online) Density dependence of the isospin chemical
potential $\mu_I$ (a) and the isospin asymmetry $\delta$ (b) in
hybrid stars with quark phase described by the 3-flavor NJL model
without isovector couplings. } \label{mudelta}
\end{figure}

The density dependence of the isospin chemical potential $\mu_I$ and the isospin asymmetry $\delta$ in hybrid stars is illustrated in Fig.~\ref{mudelta}, where $\mu_I$ turns out to be half of the charge chemical potential $\mu_c$, and the isospin asymmetry in the mixed phase is calculated from~\cite{Tor11}
\begin{equation}
\delta = \frac{(1-Y)(\rho_n-\rho_p)+Y(\rho_d-\rho_u)}{\rho_B},
\end{equation}
with $\rho_B$ defined by Eq.~(\ref{rhob}). It is seen that the
isospin chemical potential drops to as low as -115 MeV in the
hadronic phase, while it begins to increase when quarks appear.
Although the isospin chemical potential is small in pure quark
phase, it can still be large in the mixed phase. For the isospin
asymmetry, it is close to 1 at lower densities, but keeps on
decreasing with increasing baryon density. One sees from
Fig.~\ref{mudelta} that the quark system with large isospin chemical
potentials or isospin asymmetries can exist in the mixed phase of
hybrid stars. This is the existing system with the largest isospin
chemical potential we know so far, but with zero temperature and
finite net strange quark densities, different from the situation
mentioned in Sec.~\ref{diagram}.

\begin{figure}[tbh]
\includegraphics[scale=0.6]{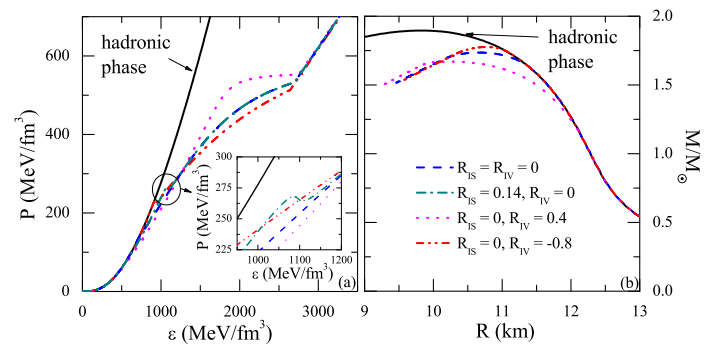}
\caption{(color online) The EOS (left) and the mass-radius relation
(right) of hybrid stars for various isovector coupling constants in
the quark phase described by the 3-flavor NJL model.  }
\label{hgisv}
\end{figure}

Figure~\ref{hgisv} displays the EOS and the mass-radius relation for
the corresponding hybrid stars from various isovector coupling
constants used in Fig.~\ref{frac}, and the result from pure hadronic
phase is also shown for comparison. Interestingly, we observe a
spinodal behavior in the EOS from $R_{IS}=0.14$ in the hadron-quark
mixed phase, as shown in the inset, exhibiting that the
corresponding hybrid star is unstable. This can be understood from
Fig.~\ref{sym} where a positive scalar-isovector coupling leads to a
significant decrease of the quark matter symmetry energy at
intermediate densities. Although not shown here, we observe that the
result from a negative $R_{IS}$ is almost the same as that without
isovector coupling. A negative $R_{IV}$ leads to the later
appearance of quarks and thus a stiffer EOS at intermediate
densities as shown in the inset, although at higher densities a
positive $R_{IV}$ somehow leads to a larger pressure. As a result, a
negative vector-isovector coupling gives the largest maximum mass of
hybrid stars (approximately 1.8 solar mass), which is still smaller
than that of the observed massive compact stars.

\section{Possible effects from breaking the strangeness sector}
\label{strangeness}

Since introducing the isovector
coupling in the 3-flavor NJL model more breaks the strangeness SU(3) symmetry than the isospin SU(2) symmetry,
here we briefly discuss the possible effects on the isospin
properties of quark matter discussed in the previous sections if we further introduce the scalar-strangeness
and vector-strangeness couplings in the NJL Lagrangian
\begin{eqnarray}
\mathcal{L}_{\rm NJL} \rightarrow \mathcal{L}_{\rm NJL} &+& G_{SS}
[(\bar{q}\lambda_8q)^2 + (\bar{q}i\gamma_5\lambda_8q)^2] \notag\\
&+& G_{SV} [(\bar{q}\gamma_\mu\lambda_8q)^2 +
(\bar{q}\gamma_5\gamma_\mu\lambda_8q)^2],
\end{eqnarray}
where $G_{SS}$ and $G_{SV}$ are the corresponding coupling
constants. As a consequence, the constituent mass, effective
chemical potential, thermodynamic potential, and energy density for
the 3-flavor NJL model are modified to
\begin{eqnarray}
M_i &\rightarrow& M_i - 2s_iG_{SS}(\sigma_u+\sigma_d-2\sigma_s),\\
\tilde{\mu}_i &\rightarrow& \tilde{\mu}_i
+2s_iG_{SV}(\rho_u+\rho_d-2\rho_s), \\
\Omega_{\textrm{NJL}} &\rightarrow& \Omega_{\textrm{NJL}} +
G_{SS}(\sigma_u+\sigma_d-2\sigma_s)^2 \notag \\
&+& G_{SV}(\rho_u+\rho_d-2\rho_s)^2,\\
\varepsilon_{\textrm{NJL}} &\rightarrow& \varepsilon_{\textrm{NJL}}
+G_{SS}(\sigma_u+\sigma_d-2\sigma_s)^2 \notag\\
&+& G_{SV}(\rho_u+\rho_d-2\rho_s)^2,
\end{eqnarray}
with $s_i=1$ for $u$ and $d$ quarks and $s_i=-2$ for $s$ quarks.

\begin{figure}[tbh]
\includegraphics[scale=0.9]{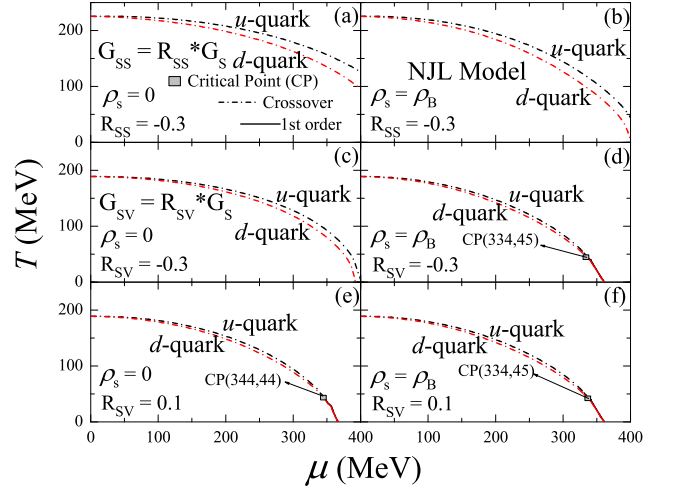}
\caption{(color online) Phase diagram in the $\mu - T$ plane for
various scalar-strangeness and vector-strangeness coupling constants
with $\rho_s=0$ (left) and $\rho_s=\rho_B$ (right) at a fixed isospin chemical potential $\mu_I=-30$ MeV from the 3-flavor
NJL model. The dash-dotted curves represent the approximate phase
boundaries of a crossover transition while the solid lines are those
of a first-order transition, with a critical point appearing in
between. } \label{strange1}
\end{figure}

Employing a fixed isospin chemical potential $\mu_I=-30$ MeV and $G_{IS}=G_{IV}=0$, we display
in Fig.~\ref{strange1} the phase diagram for various strangeness
coupling constants with $\rho_s=0$ and $\rho_s=\rho_B$
from the 3-flavor NJL model. Again we introduce $R_{SS}=G_{SS}/G_S$
and $R_{SV}=G_{SV}/G_S$ as the reduced scalar-strangeness and
vector-strangeness coupling strength, and they are treated as free parameters. As $\sigma_s$ is much larger
than $\sigma_u$ or $\sigma_d$ near the chiral phase boundary, the
constituent quark mass is mostly negative for a positive $R_{SS}$,
which is not shown here. On the other hand, for a negative $R_{SS}$
the chiral phase transition at $\mu=0$ happens at a higher
temperature, compared to the case without strangeness coupling. With finite net strange quark density $\rho_s=\rho_B$, it is seen that
generally the chiral phase boundary moves to the small-$\mu$ side, and the
isospin splitting of the chiral phase boundaries for $u$ and $d$
quarks is also slightly reduced, compared to the case with
$\rho_s=0$, as a result of the interplay between the isospin and
strangeness sectors. Again, the isospin splittings are more
considerable from the scalar-strangeness coupling than from the
vector-strangeness coupling, consistent with our findings in the
previous sections. Generally, the critical point barely exists for a
negative strangeness coupling constant, while a positive
vector-strangeness coupling constant favors the existence of the
critical point.

\begin{figure}[tbh]
\includegraphics[scale=0.6]{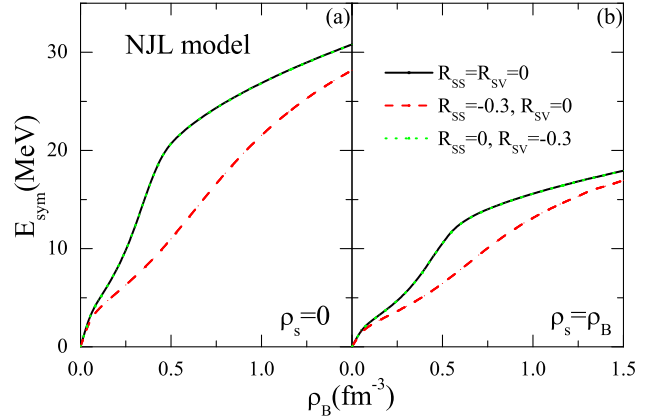}
\caption{(color online) Quark matter symmetry energy from the
3-flavor NJL model with strange quark density $\rho_s=0$ (left)
and $\rho_s=\rho_B$ (right) at zero temperature with various coupling
constants of the strangeness sector.} \label{strange2}
\end{figure}

We have also studied the effects of the strangeness coupling on the
quark matter symmetry energy, which are displayed in
Fig.~\ref{strange2}. The vector-strangeness coupling has no effect
on the quark matter symmetry energy, since the
$(\rho_u+\rho_d-2\rho_s)^2$ term is independent of the isospin
asymmetry $\delta$. Although the scalar-strangeness coupling has no
direct contribution to the potential part of the symmetry energy, it
modifies the isospin splitting of $u$ and $d$ quark constituent mass
and reduces the quark matter symmetry energy by affecting the
kinetic part. Again, the symmetry energy is reduced in the presence
of strange quarks as found in Sec.~\ref{Esym}. The EOS of isospin
asymmetric quark matter is related to the quark matter symmetry
energy, and it is found that the vector-strangeness coupling has
little effect on the hybrid star results discussed above while the
scalar-strangeness coupling slightly reduces the maximum mass of
hybrid stars.

\section{Conclusions and Outlook}
\label{summary}

In this work, we have studied the properties of isospin asymmetric
quark matter based on the 3-flavor Nambu-Jona-Lasinio model as well
as its Polyakov-loop extension with scalar-isovector and
vector-isovector couplings. Although the isospin effect on the phase
diagram has been found small with the isospin asymmetry reached in
high-energy heavy-ion collisions, considerable isospin effect is
observed at a fixed isospin chemical potential $\mu_I=-30$ MeV,
which can not be reached in heavy-ion experiments so far. The
separation of the $u$ and $d$ quark chiral phase transition is
observed with positive isovector coupling constants but is
suppressed with negative ones. The quark matter symmetry energy is
shown to decrease with increasing isovector coupling constant, and
is mostly enhanced with Polyakov-loop extension. We found that the
isospin splittings of quark condensate, constituent quark mass, and
chiral phase transition as well as the critical point are more
sensitive to the scalar-isovector coupling, while the quark matter
symmetry energy is more sensitive to the vector-isovector coupling.
A positive scalar-isovector coupling constant can lead to an
unstable isospin asymmetric quark matter and hybrid star matter. The
particle fraction as well as the equation of state in hybrid stars
depends on the isovector couplings as well. Possible effects on the
above results from further breaking of the strangeness sector among
the flavor symmetry have also been discussed.

At RHIC-BES or FAIR-CBM energies, the isospin splitting of final
hadron observables is expected to be sensitive to the isospin
dynamics of the produced quark matter. In the spirit of
Ref.~\cite{Xu14}, by comparing the experimental results of such
splitting with those from transport model simulations based on the
NJL Lagrangian used in the present work, one can in principle
extract useful information of the isospin dependence of the QCD
phase diagram and constrain the quark matter symmetry energy. In
addition, the method of calculating the QCD phase diagram used in
the present study is valid only for small isospin chemical
potentials ($|\mu_I|<m_\pi/2$). To explore the whole 3-dimensional
QCD phase diagram at larger isospin chemical potentials, one needs
to introduce an additional order parameter of pion
condensate~\cite{Son01,Kle03,Bar04,Sas10,Xia13}. Such studies will
be carried out in the future.

\begin{acknowledgments}
We thank Che Ming Ko and Feng Li for helpful discussions, and the
anonymous referee for providing useful comments and suggestions with
great patience. JX acknowledges support from the Major State Basic
Research Development Program (973 Program) of China under Contract
No. 2015CB856904 and No. 2014CB845401, the National Natural Science
Foundation of China under Grant No. 11475243 and No. 11421505, the
"100-Talent Plan" of Shanghai Institute of Applied Physics under
Grant No. Y290061011 and No. Y526011011 from the Chinese Academy of
Sciences, the Shanghai Key Laboratory of Particle Physics and
Cosmology under Grant No. 15DZ2272100, and the "Shanghai Pujiang
Program" under Grant No. 13PJ1410600. LWC acknowledges the Major
State Basic Research Development Program (973 Program) in China
under Contract No. 2013CB834405 and No. 2015CB856904, the National
Natural Science Foundation of China under Grant No. 11275125 and No.
11135011, the "Shu Guang" project supported by Shanghai Municipal
Education Commission and Shanghai Education Development Foundation,
the Program for Professor of Special Appointment (Eastern Scholar)
at Shanghai Institutions of Higher Learning, and the Science and
Technology Commission of Shanghai Municipality (11DZ2260700).
\end{acknowledgments}

\end{document}